\documentclass{emulateapj}

\newcommand{\myemail}{michal@dark-cosmology.dk}

\slugcomment{Draft version, \today}

\shorttitle{GRB-selected submillimeter galaxies}
\shortauthors{Micha{\l}owski et al.}

\begin{document}

\title{The nature of GRB-selected submillimeter galaxies: hot and young}

\author{M.~J.~Micha{\l}owski, J.~Hjorth,  J.~M.~Castro Cer\'{o}n and D.~Watson}
\affil{Dark Cosmology Centre, Niels Bohr Institute, University of Copenhagen, Juliane Maries vej 30, 2100 Copenhagen \O, Denmark}
\email{\myemail}

\begin{abstract}

We present detailed fits of the spectral energy distributions (SEDs) of four
submillimeter (submm) galaxies selected by the presence of a gamma-ray burst
(GRB) event (GRBs 980703, 000210, 000418 and 010222). These faint $\sim 3$~mJy submm
emitters at redshift $\sim 1$ are characterized by an unusual
combination of long- and short-wavelength properties, namely enhanced submm
and/or radio emission combined with optical faintness and blue colors. We
 exclude an active galactic nucleus as the source of  long-wavelength emission. From the SED fits
we conclude that the four galaxies are
young (ages $\mbox{}<2$~Gyr), highly starforming (star formation rates $\mbox{}\sim150\,{\rm M}_\odot$
yr$^{-1}$), low-mass (stellar masses $\mbox{}\sim10^{10}\,{\rm M}_\odot$) and dusty
(dust masses $\mbox{}\sim3\times10^{8}\,{\rm M}_\odot$). Their high dust temperatures
($T_d\gtrsim45$ K) indicate that GRB host galaxies are hotter, younger, and
less massive counterparts to submm-selected galaxies detected so far. Future facilities like
{\it Herschel},  JCMT/SCUBA-2 and ALMA will test this hypothesis enabling measurement of dust temperatures of  fainter GRB-selected galaxies.

\end{abstract}


\section{Introduction}

It has been claimed that submillimeter (submm) galaxies \citep[SMGs, see][for a review]{blain02} are significant contributors to the star formation history of the Universe at redshifts $z\sim2-3$ \citep{chapman05} and have built up a substantial  fraction of the present-day stellar population \citep{lilly99}. SMGs are luminous (with infrared luminosities $L_{\rm IR}\sim10^{12-14}\,L_\odot$) and cold \citep[with mean dust temperature \mbox{$T_d=36\pm7$ K,}][]{chapman05}. 
Galaxies with similar luminosities but with higher dust temperatures ($T_d>45$ K) are difficult to detect in the submm with current technology due to the fact that the peak of the infrared dust emission is shifted out of the sensitive $850\,\mu$m band towards shorter wavelengths \citep{blainT,chapman04,chapman05} in such galaxies.

At the other end of the galaxy luminosity function, the host galaxies of long-duration gamma-ray bursts \citep[GRBs, originating in the collapses of very massive stars at the end of their evolution, e.g.][]{stanek,hjorthnature}  do not have much in common with SMGs except for the fact that this type of galaxies is  also thought to contribute significantly to, or at least trace, the global star formation \citep[e.g.][]{jakobsson05,jakobsson06}.
In contrast to SMGs, GRB hosts are found to be blue, subluminous in the optical  \citep{lefloch03,fruchter06} and metal-poor \citep[][]{fynbo03}. The majority of them have not been detected at mid-infrared (MIR), submm or radio wavelengths  \citep{hanlon,lefloch,tanvir,berger,priddey06} indicating that, as a class, they are not heavily obscured or violently star forming galaxies. A low internal dust content is consistent with low extinction found in the analysis of optical afterglows \citep{stratta04,chen06,kann06} and optical spectral energy distributions (SEDs) of the host galaxies \citep{christensen04}. 

However, four GRB hosts (980703, 000210, 000418 and 010222) have been firmly detected in submm and/or radio \citep{tanvir,bergerkulkarni,berger} providing a somewhat complex picture: assuming that this emission is powered by starbursts, the derived star formation rates (SFRs) are of the order of a few hundred solar masses per year and the amount of dust in these galaxies is significant. On the other hand, they exhibit  blue colors, low extinction and low extinction-corrected optical/UV  SFRs \citep{djorgovski98,holland,sokolov01,chary02,gorosabel1, gorosabel2,galama,berger,savaglio03,christensen04,chen06,kann06},
like the majority of known GRB hosts. 
This puzzling situation was highlighted by  \citet[][their Figure 3]{berger}  --- these GRB hosts are much fainter in the optical than the prototypical Ultra Luminous Infrared Galaxy (ULIRG) Arp~220 and have much bluer spectra, but are more luminous in submm and radio. 
In fact, no
template SED model  \citep[e.g.][]{silva98,dale,dalehelou02} could give consistency with the data \citep[see also][]{michalowski06,michalowskimaster}.
Moreover, none of these hosts have been detected at $24\,\mu$m \citep[Le Floc'h, private communication]{lefloch,castroceron06}. 

The location of the GRB events within their hosts adds further  complexity to this picture. As found by \citet{fruchter06}, GRBs trace the location of the brightest rest-UV parts of their hosts \citep[see also][]{bloom02}. If the majority of the star formation in the hosts of GRBs 980703, 000210, 000418 and 010222 was hidden by dust, then they should preferentially be found in obscured (UV-dim) parts of their hosts (as long as GRBs trace star formation), contrary to observations.

In this paper we  investigate this seeming discrepancy between short- and long-wavelength data through stellar population model SED fitting.  In particular we discuss the possibility that these submm-bright GRB hosts may represent the long-sought hotter (and less luminous) counterparts of SMGs. In Section~\ref{sec:model} we describe the model and the fitting procedure, in Section~\ref{sec:results} we show the results,  discuss their implication in Section~\ref{sec:discussion}, and finally in Section~\ref{sec:conclusion} conclude this work.
We use a cosmological model with $H_0=70$ km s$^{-1}$ Mpc$^{-1}$, 
$\Omega_\Lambda=0.7$ and $\Omega_M=0.3$.

\section{GRASIL SED Modeling}
\label{sec:model}

In order to model the SEDs of GRB hosts we used the GRASIL\footnote{\url{http://web.pd.astro.it/granato/grasil/grasil.html}}
software developed by \citet{silva98}.
It is a numerical code that calculates the spectrum of
a galaxy by means of a two-dimensional radiative transfer method, applied to photons produced by a stellar population,
and reprocessed by dust. This model is self-consistent
in that it fulfills the principle of energy conservation between the energy absorbed by dust in the UV/optical wavelengths
and the energy re-emitted in the infrared. Two extinction media are implemented,   dense star forming molecular clouds (MCs, applied only to the youngest stellar population) and  diffuse cirrus.   The dust is composed of small  grains (not in thermal equilibrium with radiation and hence fluctuating in temperature), big grains (silicates and graphites) and polycyclic aromatic
hydrocarbon (PAH) molecules. The emission of grains with given size is assumed to be a grey-body, so the composite spectrum is a sum of all grey-bodies with different temperatures. A galaxy is assumed to be an axially symmetric system. Different geometries of the stellar and diffuse dust distributions are allowed but were not used here. The SFR is assumed to be proportional to the available gas content following the Schmidt-law. On top of this smooth SFR history a violent starburst epoch  is added. Star formation is unevenly distributed throughout the galaxy in MCs. The SFR given as a GRASIL output is the sum of the SFRs of each MC.

We gathered photometric data from the literature at optical \citep{sokolov01,vreeswijk,gorosabel1,gorosabel2,galama}, infrared \citep{castroceron06,
lefloch}, submm \citep{tanvir,berger}  and radio \citep{bergerkulkarni,berger,frail, sagarstalin01} wavelengths. For the host of  GRB 000210 we performed the photometry on the archival {\it Spitzer} images \citep[see][for a decription of the procedure]{castroceron06,castroceron07} and obtained the following fluxes: at $4.5\,\mu$m: $3.3\pm2.1\,\mu$Jy; at $8.0\,\mu$m: $15.0\pm5.1\,\mu$Jy and at $24\,\mu$m: $<31.5\,\mu$Jy ($3\sigma$).

We performed  SED modeling investigating a wide range of the following GRASIL parameters:
age of the galaxy (defined as the time since the beginning of its evolution when the stellar population started to build up), dust-to-gas ratio, and mass of gas converted into stars during the current starburst episode (lasting  $50$ Myr). We used a clearing time for MCs, $t_{\rm esc}=50$ Myr \citep[see][for discussion of this parameter]{panuzzo07}. 

\section{Results}
\label{sec:results}
\begin{figure*}
\begin{center}
\plotone{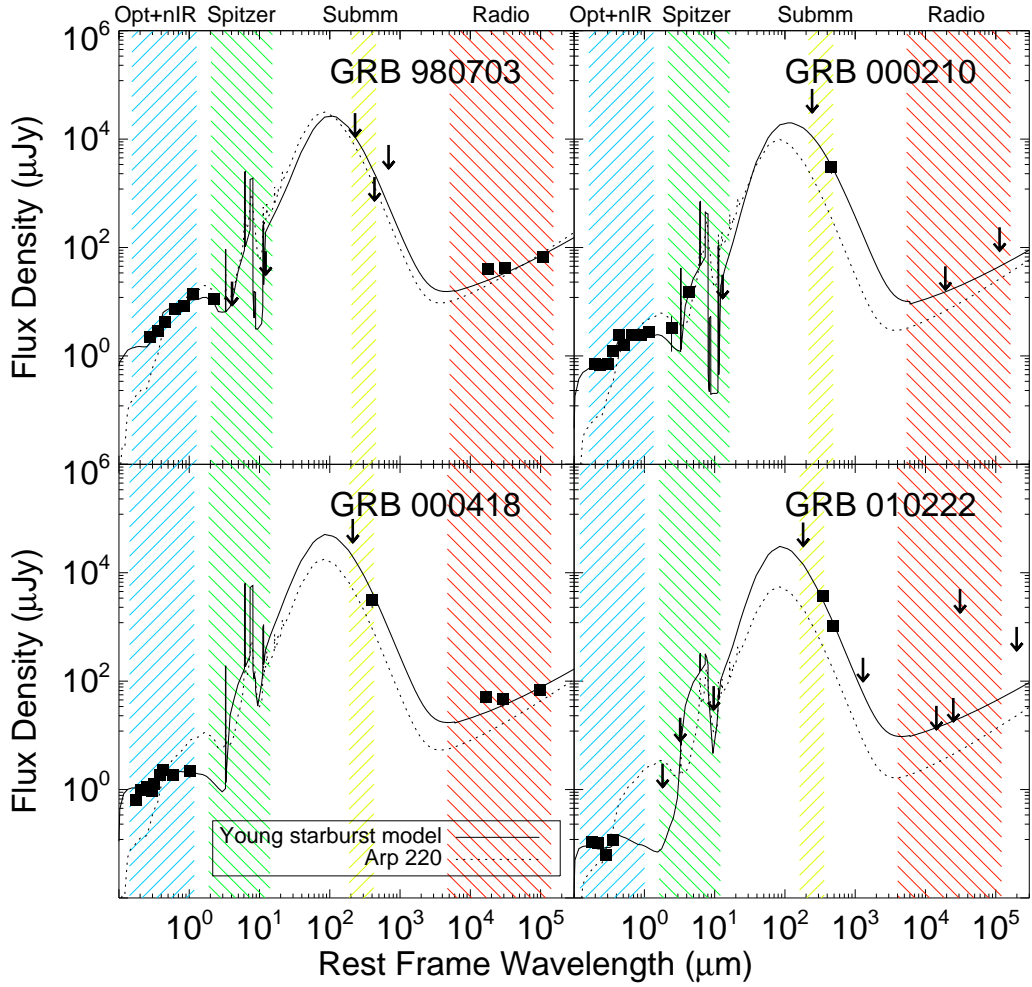}
\end{center}
\caption{SEDs of GRB hosts. {\it Solid lines}: the young starburst galaxy models calculated using GRASIL and consistent with the data. {\it Dotted lines}: Arp 220 model \citep[from][]{silva98}.
{\it Squares}: detections with errors, in most cases, smaller than the size of the symbols. {\it Arrows}: $3\sigma$ upper limits (values marked at the base).
Hashed columns mark the wavelengths corresponding
to optical/near-infrared filters, MIR {\it Spitzer} filters, SCUBA submm bands and the radio domain. It is apparent that the ULIRG Arp 220 model underpredicts the radio and submm fluxes of GRB hosts while fitted to the optical data. The optical color of this model is also too red compared to GRB hosts. In contrast, the emission of young starbursts is  blue in optical and strong in submm/radio at the same time. In the case where we lack {\it Spitzer} data the MIR part of the SED is unconstrained due to the possible absorption or emission of PAH and silicate features.}
\label{fig:sed}
\end{figure*}
{
\begin{table*}
\caption{GRASIL parameters yielding SEDs consistent with the data and characteristics of the galaxies derived from the SED modeling.}
\begin{center}
\begin{tabular}{c r@{.}l | r@{}l r@{}l r@{}l       |       r@{.}l c r@{.}l cccc}
\hline\hline
\multicolumn{1}{c}{GRB}& \multicolumn{2}{c}{$z$}& \multicolumn{2}{|c}{Age} & \multicolumn{2}{c}{dust/gas} &  \multicolumn{2}{c|}{$M_{\rm burst}$}& \multicolumn{2}{c}{$L_{\rm IR}$} &			SFR &	\multicolumn{2}{c}{$M_*$} & $M_d$ & $T_d$ & $A_{\rm 1\mu m}^{\rm MC}$ & $A_V^{\rm av}$\\
host	& \multicolumn{2}{c}{ } & \multicolumn{2}{|c}{(Gyr)} & \multicolumn{2}{c}{($10^{-2}$)}&  \multicolumn{2}{c|}{($10^9\,{\rm M}_\odot$)} &\multicolumn{2}{c}{($10^{12}L_\odot$)} &			(${\rm M}_\odot$ yr$^{-1}$)&			\multicolumn{2}{c}{($10^{9}\,{\rm M}_\odot$)} & ($10^9\,{\rm M}_\odot$) & (K) & & \\
	(1) & \multicolumn{2}{c}{(2)} & \multicolumn{2}{|c}{(3)} & \multicolumn{2}{c}{(4)} & \multicolumn{2}{c|}{(5)} & \multicolumn{2}{c}{(6)} & (7) & \multicolumn{2}{c}{(8)} & (9) & (10) & (11) &(12)\\
\hline
\object[GRB 980703]{980703}& 	0&97&  2& & \mbox{\hspace{1.5em}}0&.3 &  \mbox{\hspace{1em}}2&.3&	\mbox{\hspace{1.5em}}1&9& 212&		\mbox{\hspace{1em}}21&0 & 0.15 & 44 & 54 & 0.42\\
\object[GRB 000210]{000210}& 0&85& 0&.19 & 2&.4 &  1&.2& 0&9&		179&		10&1 & 0.33 & 45 & 27 & 1.01\\
\object[GRB 000418]{000418}&  1&12&  0&.14&    1&  & 5&.0 &	4&6& 	288&		18&7 & 0.82 & 50 & 31 & 0.70 \\
\object[GRB 010222]{010222}&  1&48& 0&.09 & 0&.7 &  5&.9&	4&3&	278&		13&5 & 0.32 & 51& 25 & 0.58\\
\hline
\object{Arp 220}& 0&018 &13& & 1& &  25& & 1&16 & 580 &  153&2     & 0.30 & 55 & 63 & 0.59\\ 
\object{NGC 6946}& 0&00016 & 13& & 1& &  - & & 0&05 & 6 &  104&1  & 0.10   & 29 & 36 & 0.06\\ 
\hline
\end{tabular}
\label{tab:grasilres}
\label{tab:grasilparam}
\tablecomments{(1)~GRB name, (2)~redshifts from \citet{djorgovski98}, \citet{piro}, \citet{bloom03b}, \citet{jha01}, \citet{arp220} and \citet{ngc6946}, (3)~age of the galaxy defined as the time since the beginning of its evolution, (4)~dust-to-gas ratio, 
(5)~mass of gas converted into stars during starburst, (6)~total $8-1000\,\mu$m infrared luminosity (7)~total star formation rate for $0.15-120\,{\rm M}_\odot$ stars averaged over the last 50 Myr, (8)~stellar mass, (9)~dust mass, (10)~dust temperature, (11)~extinction in molecular cloud (MC) at 1 $\mu$m measured from its center, (12)~average extinction of stars outside MCs at 0.55 $\mu$m. Values corresponding to
Arp 220 (ULIRG) and NGC 6946 (spiral) are given for comparison.}
\end{center}
\end{table*}
}

The best fits\footnote{The SED fits can be downloaded from\\ \protect\url{http://archive.dark-cosmology.dk}} 
are shown in Figure~\ref{fig:sed} and the results for each parameter listed in Table~\ref{tab:grasilparam}
together with  several properties of the galaxies derived from the SEDs. SFRs, stellar masses, dust masses and extinction in the MCs are given as output from GRASIL. Infrared luminosities were obtained by integrating  the SEDs over the range of $8-1000\,\mu$m. Dust temperatures were estimated by fitting a grey-body curve to the part of the SEDs near the dust peak \citep[$\sim100\,\mu$m, e.g.][]{yun}. Finally, the average extinction outside MCs was calculated as: $A_V=2.5\log$($V$-band starlight extinguished by MCs only / $V$-band starlight observed); see \citet{silva98}. This parameter describes the extinction averaged throughout the galaxy as opposed to the line-of-sight extinction derived from optical GRB afterglows.

We checked the consistency of our results with those reported in the literature 
\citep[see Appendix \ref{sec:comparison}]{berger01,berger,bjornsson02,castroceron06,castroceron07,chen06,christensen04,galama,gorosabel1,gorosabel2,lefloch,michalowskimaster,michalowski06,sokolov01,stratta04,takagi04}.
We  generally found good agreement except for the following points. Our estimate of the age for GRB 980703 (which has the most significant old stellar component) is considerably larger than that derived by \citet{christensen04}. The discrepancy arises because  we calculated the time
from the beginning of the galaxy evolution, not the beginning of the starburst.
Our SFRs for GRB 980703 and 010222 are inconsistent with the reported upper limits of \citet{lefloch}. However our analysis, based on the full SED rather than only $24\,\mu$m datapoints, seems to be more reliable as admitted by \citet{lefloch} who assigned a factor of $\gtrsim5$ error to their estimates \citep[see also the results of][based on the same data]{castroceron06}. 
\citet{berger} obtained systematically higher SFRs by a factor of $2$--$5$ 
based on submm alone. This may be the effect of uncertainty of the dust properties  (we could reproduce the results of \citet{berger} using \citet{yun}, but when we used a  temperature $10$ K higher and an  emissivity   index $0.35$ lower than \citet{berger} did, we obtained values consistent with our results reported here) as well as of the contribution of older stellar populations to the submm fluxes \citep[see for example the discussion in][]{vlahakis06}, that leads to an overestimation of SFRs using only submm data.
For the hosts of GRB 000210, 000418 and 010222, \citet{takagi04} predicted much higher SFRs, of the order of a few thousand \mbox{${\rm M}_\odot$ yr$^{-1}$} and stellar masses of \mbox{$\sim10^{11}\,{\rm M}_\odot$}. Such a high SFR is unlikely to be necessary to explain the submm emission of GRB hosts and hence the accumulated stellar masses seem to be too high as well.
We derived much higher extinction for the host of GRB 000210 than \citet{christensen04}, but our model predicts that the majority of the extinction in the \mbox{$V$-band} has a grey nature, which would be undetectable in any reddening measurements (we found $E(B-V)=0.07$, which is consistent with their results if one assumes the Galactic extinction curve slope $R_V=3.1$). Moreover, it is possible that a GRB event  destroys the dust along the line of sight, so study of an afterglow  results in low dust content.

The determination of the infrared luminosity suffers from systematic uncertainties depending on the choice of the SED template. Our approach of using all the optical, submm and radio data to constrain the shape of the SED results in a moderate systematic error in luminosity. Using the templates of \citet{dale} and \citet{dalehelou02} fitted to long-wavelength data we obtained values only 30\% lower. Similar analysis on a bigger sample of galaxies led  \citet{bell07} to the conclusion that the systematic uncertainty of infrared luminosity is usually less than a factor of $\sim2$. 
The arbitrary choice of the \citet{salpeter} IMF with the cutoffs of $0.15$ and $120\,{\rm M}_\odot$ introduces a systematic error of a factor of $\sim2$ in the determination of the stellar masses and SFRs \citep{erb06}. \citet{bell07} have also found that random errors in stellar mass are less than a factor of $\sim2$. Dust mass estimates are uncertain up to a factor of a few  
\citep{silva98}.
Estimates of dust temperatures based on submm and radio alone have uncertainties of $\sim10$ K \citep{chapman03}.

\section{Discussion}
\label{sec:discussion}

\subsection{Solving the puzzle}
\label{sec:puzzle}

The key property of GRB hosts that explains their blue colors and enhanced submm/radio emission is their young ages (Table~\ref{tab:grasilparam}).  On one hand the majority of the stars still reside in dense MCs, so a significant part of the energy is absorbed and re-emitted. This increases the dust emission. On the other hand there are lots of young, hot, blue stars in such a galaxy, because they have not finished their lives yet. Hence the total optical spectrum of the galaxy is blue.
GRBs may indeed reside in or close to MCs \citep[possibly causing hydrogen ionization and dust sublimation along the line of sight,][]{watson07} 
and it was  found that gas column densities
derived from {\it X}-ray afterglows in a sample of 8 GRBs (including GRB 980703, discussed here) were in the range
corresponding to the column densities of giant MCs in the Milky Way \citep{galamawijers01}.
A similar conclusion for high-redshift GRBs was recently drawn by \citet{jakobsson06b} by means of modeling Ly$\alpha$ absorption features \citep[see also][for a discussion of similar results]{castrotirado99,hjorth03,savaglio03,stratta04,vreeswijk04,chen05,chen06,watson06,campana07,prochaska07b,prochaska07, ruizvelasco07}. 

Since the stars formed during the starburst do not dominate the stellar mass of the GRB hosts discussed here (only from 10 to 40\% of these masses have been formed during the on-going starburst epoch --- compare columns  5 and 8 of  Table  \ref{tab:grasilres}) and are still embedded in MCs providing strong extinction \citep[so-called age-dependent extinction, see][]{panuzzo07}, it is apparent that the optical/UV light is dominated by somewhat older stars, which are already outside MCs and suffer only moderate extinction \citep[column 12 in Table~\ref{tab:grasilres}, see][for an example of a totally obscured, metal-poor star-forming region]{plante02}. This is  why optical measurements of extinction resulted in low values  suggesting low dust content 
 whereas enhanced submm emission is consistent with being emitted by dusty galaxies. In light of this, we provide support for the hypothesis of \citet{gorosabel1} claiming that the optical and submm emission of the GRB 000210 host are dominated by different populations of stars \citep[the same explanation was proposed by][for a sample of dusty ULIRGs with UV colors bluer than expected for starburst galaxies]{goldader02}.

The location of GRBs in the brightest UV parts of the galaxy \citep{fruchter06} is also consistent with our model. Although the GRBs discussed here trace regions of obscured rather than non-obscured star formation (because the majority of star formation is obscured and under the assumption that GRBs trace star formation), these regions are not spatially distinct in the galaxy. Regions of obscured star formation evolve into non-obscured regions by destroying the MCs without changing their location. Unless individual MCs can be resolved, obscured star formation within them and less-obscured star formation on their outskirts cannot be spatially separated.

Our results are based on the assumption that the detected submm/radio sources are indeed related to GRB hosts. It is however possible that the emission comes from unrelated sources falling into the coarse beam of SCUBA ($15''$) as noted by \citet{smith01,smith05}, \citet{gorosabel1} and \citet{lefloch}. The most suspicious case is the GRB 010222 host which is undetected at $24\,\mu$m, but accompanied by several spatially close MIR-bright galaxies that could dominate the submm emission \citep{lefloch}. On the other hand, the nondetections of GRB hosts in the MIR are easily explained by silicate absorption features and the steep infrared spectrum of the galaxies (see Figure~\ref{fig:sed}).  The situation is less severe for GRB 980703 and GRB 000418, both detected by VLA in the radio, for which accurate astrometry decreases the chance of confusion.

\subsection{The nature of the GRB hosts}
\label{sec:nature}

From the results presented in Table  \ref{tab:grasilres} a common characteristic of submm GRB hosts emerges (although our sample is too small to draw a very robust conclusion).
For all four galaxies we obtained young ages ($\lesssim2$ Gyr) and relatively low stellar masses ($\lesssim2\times10^{10}\,{\rm M}_\odot$). 
These are usual properties  among GRB hosts \citep{christensen04,sollerman05,castroceron06,savaglio06}. Galaxies with such stellar mass are believed to dominate the star formation of the Universe at redshifts $z\lesssim1$ \citep{zheng07,bell07}.

What is special about the four galaxies discussed here is that they are highly star forming \citep[as opposed to the other known GRB hosts, see][]{berger,lefloch,tanvir} and have high dust content. Hence they can be  representative only of the bright end of the infrared luminosity function of GRB hosts. More sensitive observations in submm (by
\mbox{JCMT/SCUBA-2} and ALMA) and far-infrared (FIR, by {\it Herschel}) are therefore necessary to detect fainter hosts and test this hypothesis.

The optical afterglow of GRB 000418 was exceptionally red \citep{klose00}  while no optical afterglow was detected for GRB 000210  in spite of deep searches \citep{piro} \citep[its {\it X}-ray luminosity was high enough to place it at the borderline of the category of dark GRBs,][]{jakobsson04,rol05}, both hinting at significant obscuration in the hosts. However the ``darkness'' of GRBs cannot easily  be linked with obscuration in the hosts, because the remaining three members of our submm/radio sample were associated with bright optical afterglows \citep{berger01,bloom98,bjornsson02,castrotirado99, galama,holland,klose00,vreeswijk}. Moreover, \citet{barnard03} did not detect submm emission from the hosts of three GRBs classified as dark. This issue should be addressed by targeting a more significant sample of dark GRBs in the submm and radio.
\begin{figure*}
\plotone{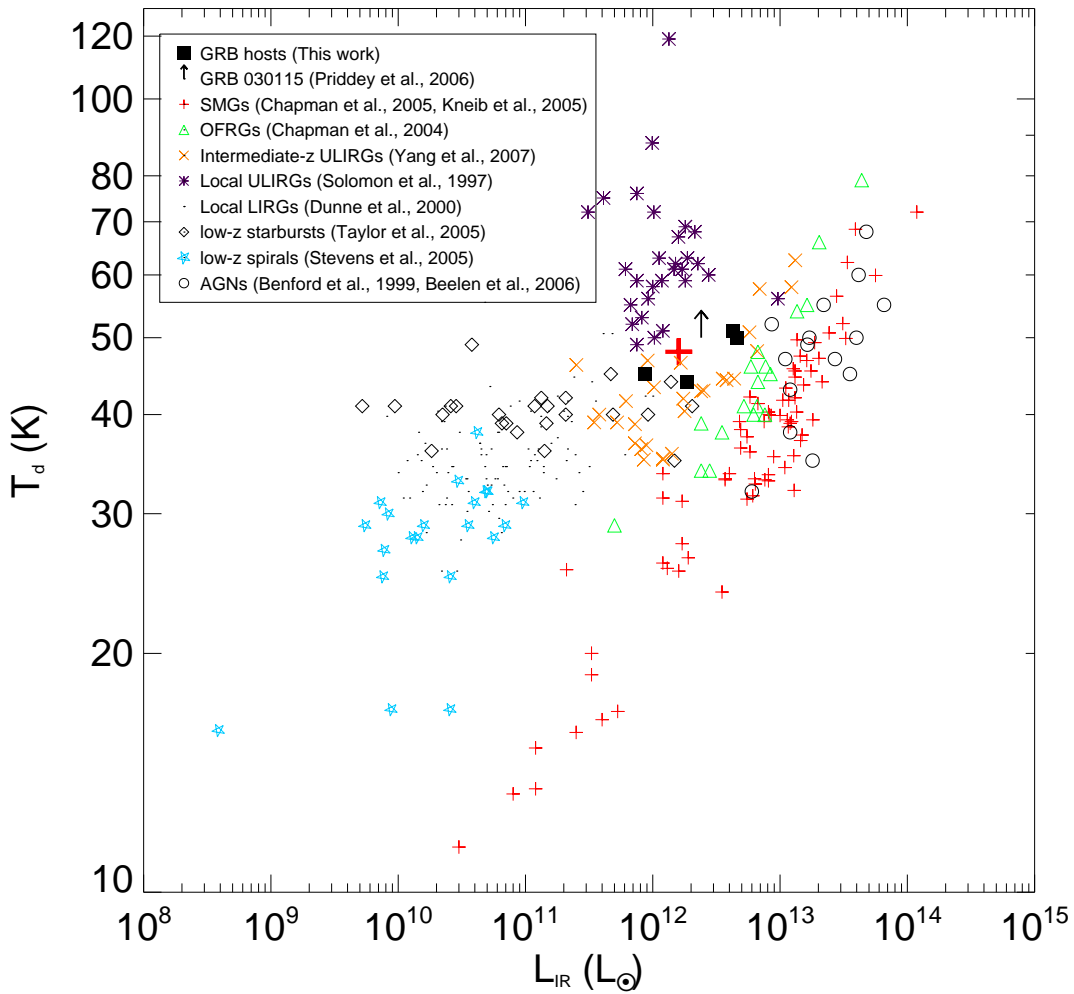}
\caption{ Dust temperature as a function of infrared ($8-1000\,\mu$m) luminosity. GRB hosts discussed here ({\it filled squares}) and GRB 030115 \citep[{\it arrow} indicating lower limit on temperature,][]{priddey06} are compared with submm galaxies \citep[{\it pluses},][large symbol denotes a hot lensed galaxy found behind a cluster]{chapman05,kneib05}, optically faint radio galaxies \citep[{\it triangles}, note that symbols indicate lower limits on temperature,][]{chapman04}, \mbox{intermediate-$z$} ULIRGs \citep[{\it crosses},][]{yang07}, local ULIRGs \citep[{\it asterisks},][]{solomon97}, local LIRGs \citep[{\it dots},][]{dunne00}, low-$z$ starburst galaxies \citep[{\it diamonds},][]{taylor}, low-$z$ spirals and cirrus galaxies \citep[{\it stars},][]{stevens05,taylor} and  millimeter-selected radio-quite AGNs \citep[{\it circles},][]{benford99,beelen06}. GRB hosts seem to reflect the properties of intermediate-$z$ ULIRGs, the bright end of starburst galaxies and optically faint radio galaxies --- the candidates for hotter counterparts of SMGs. Moreover, the hot, faint SMG found behind the cluster ({\it large plus}) falls in the same region as GRB hosts.}
\label{fig:T_L}
\end{figure*}
\subsection{Rejection of an AGN contribution}

In general, GRB hosts are typically starburst galaxies and this selection makes the presence of an AGN component quite unlikely. Here we present additional indications that the long-wavelength emission from GRB hosts discussed here is not dominated by AGNs.

In order to assess the probability that there is an AGN component in our sample of GRB hosts we compared their FIR and radio luminosities using the $q$ coefficient of \citet{helou85}: $q=\log(\mbox{FIR}/3.75\cdot10^{12}/S_{1.4})$. The FIR luminosity was integrated in the range $42.5$--$122.5\,\mu$m. The rest $1.4$~GHz luminosity ($S_{1.4}$) was calculated from the observed $1.4$~GHz (or 4.86~GHz for GRB 010222) flux assuming a steep radio slope $\alpha=-0.75$ to obtain a robust lower limit on $q$. (Assuming a shallower slope, the expected rest $1.4$ GHz radio luminosity would be even lower). The resulting $q$ values are  $2.28$, $>1.93$, $2.41$ and $>2.06$ for GRBs 980703, 000210, 000418 and 010222,  respectively. It is known that starburst galaxies follow the so-called FIR--radio correlation with a mean $q=2.21\pm0.14$ \citep{helou85} or $2.3\pm0.2$ \citep{condon}. Hence the GRB hosts are consistent with being starburst-dominated galaxies. Emission dominated by radio-loud AGN would show $q<2$ \citep{yun01,chanial07,yang07} which would only be the case for GRB 000210 if the actual radio flux is just below the $3\sigma$ upper limit. We therefore conclude that the FIR-radio correlation shows that the emission from GRB hosts is probably not dominated by radio-loud AGNs. Even if a non-dominating AGN is present its contribution to the radio and submm emission is insignificant since the hosts fulfill the FIR-radio correlation. It would be rather unlikely that AGN-domianted emission is coincidentally consistent with tight FIR--radio correlation.

This is supported by several other diagnostics. 
\citet{bergerkulkarni} detected no variability in the radio flux of GRB 980703 host over 650 days contrary to what would be expected if an AGN dominates its radio emission. Moreover, GRB hosts have optical spectra typical for starbursts, not AGNs \citep[i.e.~no high-ionization AGN lines
have been found,][]
{djorgovski98, bergerkulkarni,bloom03b,prochaska04}. This excludes non-obscured AGNs in our sample.
Moreover, GRB hosts  are dwarf galaxies and the fraction of AGNs (without optical high ionization lines) in a galaxy sample with stellar mass of $M_*\sim10^{10}\,{\rm M}_\odot$ is negligible ($<0.01$\% from Figure~2 of \citeauthor{best05} \citeyear{best05}; see also \citeauthor{woo05} \citeyear{woo05}). AGNs are also more luminous than GRB hosts as is shown in Figure~\ref{fig:T_L}. 
In summary, AGNs cannot dominate the emission of the GRB hosts discussed here unless they are atypically small, obscured and radio-quiet. Since submm- and radio-faint hosts are even less probably connected with AGN activity (because it would require even smaller AGNs) we conclude that GRB hosts in general are unlikely powered by AGNs.

\subsection{The general picture of dust properties}
\label{sec:T_L}

In Figure~\ref{fig:T_L} we compare the total infrared luminosities  and dust temperatures of GRB hosts with well-studied galaxies both local and at high-$z$. We included the four hosts studied here as well as the host of GRB 030115 which has an upper limit to the temperature of 50 K. This limit was derived by \citet{priddey06} using SED modeling of optical and near-infrared data, which led to the estimation of the infrared luminosity  via the UV-slope method. This, in turn, allowed them to exclude a low dust temperature, because cold dust would be inconsistent with the non-detection in the submm.

GRB hosts seem to overlap with intermediate-$z$  ULIRGs from \citet{yang07}. This is not very surprising since both galaxy classes 
have ULIRG characteristics. GRB hosts are however much more distant. As opposed to a mean redshift of $0.37$ for intermediate-$z$  ULIRGs,  our  sample has a  mean redshift of $1.1$ compared to $2.8$ for GRBs in general \citep{jakobsson06}, $2.2$ for SMGs \citep{chapman05} and $2.1$ for optically faint radio galaxies \citep[OFRGs,][]{chapman04}.

The GRB hosts discussed here occupy the same region of Figure~\ref{fig:T_L} as the brightest starbursts from \citet{taylor}. Moreover, dust masses of GRB hosts (Table~\ref{tab:grasilres}) are close to the upper boundary for starburst galaxies derived by these authors. 

As noted above, the GRB hosts discussed here must form the bright end of the infrared luminosity function of GRB hosts. The remaining members of the sample are still undetected at long wavelengths making it impossible to measure luminosities and dust temperatures. One can speculate from Figure~\ref{fig:T_L} that they may follow the ``starburst'' sequence having similar temperatures ($T\sim40-50$ K) and infrared luminosities in the range $10^9-10^{12}\,L_\odot$. {\it Herschel} should be able to detect these sources in the FIR. There is also the possibility that  we can detect even brighter (but rare) GRB hosts --- the counterparts of the brightest ULIRGs of \citet{yang07} with luminosities $\sim10^{13}\,L_\odot$ and temperatures $\sim60$ K.  These should be bright enough to be searched  for by sensitive  submm instruments (\mbox{JCMT/SCUBA-2}, ALMA).

It is known that GRB hosts are much bluer than massive, star forming SMGs \citep[compare][]{christensen04,lefloch03,berger,smail04}. From Figure~\ref{fig:T_L} it seems that they are also hotter than SMGs with the same luminosity (or dimmer than SMGs with the same temperature). This gives a hint that GRB events may pinpoint a population of ULIRGs at high redshifts with dust hotter than in typical SMGs. 
The search for such galaxies is important because they likely contribute to the star formation history at the same level as SMGs.
High dust temperatures of GRB hosts were hypothesized by \citet{barnard03} and \citet{tanvir} as a possible way to explain their faintness in submm. Here we provide  evidence that this is the case. Moreover, \citet{trentham02} suggested that GRB hosts may be low-luminosity SMGs.
Indeed, very faint sub-mJy SMGs magnified by clusters of galaxies are found to be hotter than those found in blind surveys (limited by confusion to $\sim 2$ mJy): the $z\sim2.5$ SMG behind the cluster A 2218 found by \citet{kneib05} has very similar dust properties to the GRB hosts discussed here ($T_d\sim50$, $L_{\rm IR}\sim10^{12}\,L_\odot$).

Hotter dust temperatures  indicate that the star formation sites in GRB hosts are more compact than those in SMGs \citep{chanial07,yang07}. It is consistent with the fact that GRB hosts have higher specific SFRs (per unit mass) than SMGs \citep{castroceron06}.
 The blue optical colors of GRB hosts compared to SMGs can also be explained by the compactness of the former galaxies.  If they are compact then the stellar population suffers strong extinction. This would lead to redder colors, but it is likely (see Section \ref{sec:puzzle})  that this extinction is so strong that  very young stars are totally obscured, so optical light is dominated by relatively less
obscured stars outside star forming regions leading to blue colors.

We note that the majority of the galaxies shown in Figure~\ref{fig:T_L}  also have higher dust temperatures compared to SMGs. However, all of them are local galaxies, so cannot be considered as counterparts of high-redshift SMGs and their submm emission has been detected only because of their proximity.

GRB hosts may be consistent with a population of OFRGs having similar infrared luminosities and (likely) temperatures. 
Although the majority of OFRGs lie at $z\sim2$ \citep{chapman04}, some of them are within the redshift range of the GRB hosts discussed here.
OFRGs have been suggested to be hotter counterparts of SMGs \citep{chapman04}, so the same may be true for GRB hosts.

Indeed SMG samples are clearly biased against the high-$T_d$--low-$L$ galaxies \citep[upper-left part of Figure~\ref{fig:T_L}, see][]{chapman05}. The lack of SMGs in low-$T_d$--high-$L$ (lower-right corner) is probably real, because such sources would be detected if present \citep[see Figure 2 of][for a discussion of selection effects]{blainT}. Similarly,  the lack of very luminous sources with $L>10^{14}\,L_\odot$ is probably real. If such powerful sources exist, they are very rare and do not contribute to the presented sample.
Taking into account all the galaxies shown in Figure~\ref{fig:T_L}, the lack of high-$T_d$--low-$L$ galaxies is probably not a selection effect (at low redshift) because the hotter counterparts of $10^{10}\,L_\odot$ galaxies should be easily detected in MIR and FIR.

Therefore the apparent trend (also seen in our GRB host sample) that more luminous galaxies have higher dust temperatures is a real effect called the luminosity-temperature relation \citep{soifer87,chanial07,yang07}. The spread has been interpreted as variation in the size of a star forming region \citep{chanial07} or in the dust emissivity index and amount of dust in each galaxy  \citep{yang07}. Galaxies with large dust content tend to occupy the lower-right corner of Figure~\ref{fig:T_L} whereas those with low dust content occupy the upper-left corner. The GRB hosts discussed here with $M_d\sim10^8\,{\rm M}_\odot$ are placed near the average of all the galaxies shown in Figure~\ref{fig:T_L}. The rest of the population is probably aligned along the diagonal of Figure~\ref{fig:T_L} fulfilling the luminosity--temperature relation. It is expected that some of the IR-faint GRB hosts have much lower dust content so are possibly located in the lower-$L$---higher-$T_d$ regime.

\section{Conclusions}
\label{sec:conclusion}

The short- and long-wavelength properties of the host galaxies of GRBs 980703, 000210, 000418 and 010222 can be linked assuming that they are very young and powerful starbursts.  We conclude that they are galaxies with  the highest star formation rates  among known GRB hosts, but their optical properties, starburst nature, stellar masses and ages are not distinctive. We also found that AGNs are probably not responsible for boosting their long-wavelength emission. 

We have shown that  GRB host galaxies at cosmological redshifts may constitute a population of hot submillimeter galaxies. This hypothesis makes GRB hosts of special interest, placing them in the same category as optically faint radio galaxies, and should be confirmed by  future sensitive long-wavelength  observations. Future instruments ({\it Herschel},  
JCMT/SCUBA-2 and ALMA) will be able to build up a statistically significant sample of GRB hosts with accurately measured infrared luminosities and temperatures.

\acknowledgments
We wish to thank J.~Baradziej, D.~Alexander, B.~Cavanagh, F.~Economou, J.~Fynbo, T.~Greve, T.~Jenness, E.~Le Floc'h, A.~Levan, P.~Natarajan, C.~Skovmand for help, discussion and comments and L.~Silva for making the GRASIL code available.
The Dark Cosmology Centre is funded by the Danish National Research Foundation. MJM would like to acknowledge support from The Faculty of Science, University of Copenhagen.
JMCC gratefully acknowledges support from the Instrumentcenter for Dansk Astrofysik and the Niels Bohr Institutet's International PhD School of Excellence.

\appendix

\section{Comparison of our results with the literature}
\label{sec:comparison}

In the following appendix we compare our results with those found by other authors applying different methods, to show that our modeling in most cases gives consistent values, but also provides additional galaxy properties that cannot be inferred from previous approaches. We  stress that some inconsistency with methods based on optical/UV is expected because optical/UV light traces only a minor (i.e.~unobscured) portion of the bolometric luminosity of the galaxies. Some authors used different IMFs, but the necessary correction to total SFRs and stellar masses is not larger than 15\%, and does not affect the conclusions of the comparison.

\subsection{Ages}

Based on galaxy SEDs fitted to the optical data only, \citet{gorosabel1,gorosabel2} and \citet{christensen04}
derived the ages of the starbursts in the hosts of GRBs 980703, 000210 and 000418.
Our estimates are considerably larger only for GRB 980703. The discrepancy is because  we calculated the time
from the beginning of the galaxy evolution, not the beginning of the starburst.
\citet{sokolov01} derived ages of both old stellar populations and starbursts.
Our estimation for GRB 980703 agrees
with the age of the old component, which is conceptually closer to our definition
of the galaxy age. 
Ages derived by \citet{takagi04} for GRB 000210, 000418 and 010222 agree with our results within a factor of a few.

\subsection{Star formation rates}

See \citet{michalowski06} for details of the comparison between SFR estimates.
Since the starburst is still hidden in MCs and its optical light is extinguished, SFRs derived from optical indicators \citep{christensen04,gorosabel1, gorosabel2,berger} are two orders of magnitude lower than our estimates.
Our results are consistent with radio-derived SFRs \citep{berger}. This is because the calibration of SFR to
radio flux requires the prior assumption of only two parameters  \citep[a normalization factor and a spectral index, see][]{yun},
which are relatively well constrained. Our values also agree with the upper limits derived by \citet{castroceron06}
using the template of Arp 220.
Finally, \citet{berger} obtained systematically higher SFRs by a factor of $2-5$ 
based on submm alone.  We have checked that the SFRs derived from our SED models using the total infrared emission \citep{kennicutt} are consistent with our values derived here \citep[see][]{michalowski06}.

\subsection{Stellar masses}
\label{sec:stellarmass}

Our results for the GRB 980703 host agree with the  stellar mass derived by \citet{castroceron06} and with  both stellar and burst masses derived by \citet{sokolov01}. The  stellar mass derived by \citet{castroceron07} for GRB 000210 is also in agreement with our value.

\subsection{Dust properties}

 Our value of the dust mass for GRB 980703 host agrees within a factor of $1.5$ with the one derived by \citet{castroceron06}.
 Dust masses derived by \citet{takagi04} for GRB 000210, 000418 and 010222 agree with our results within a factor of a few. 
We have checked that for these three hosts, values of dust masses computed directly from submm fluxes \citep[using][]{hildebrand,taylor} agree with those reported here \citep[see][]{michalowskimaster}.

The near-infrared extinction derived for MCs (column 11 in Table~\ref{tab:grasilres}) is within the typical values found in observations of compact star forming regions \citep[e.g.][]{scoville98,plante02,hunt05} and numerical simulations \citep[e.g.][]{indebetouw06,goicoechea07,panuzzo07}. 
Our values of average extinction outside MCs, $A_V$ (column 12 in Table~\ref{tab:grasilres}), are consistent with those derived from optical host SED modeling by \citet[][for 980703]{sokolov01}  and \citet[][for 980703, 000210, 000418]{christensen04} and from optical afterglow modeling by \citet[][for 000418]{berger01}, \citet[][for 010222]{bjornsson02}, \citet[][for 980703, 010222]{stratta04} and \citet[][for 980703]{chen06},  except for the host of GRB 000210 for which we predict higher extinction, but with grey nature undetectable in  any reddening measurements 
 To the best of our knowledge we present the first $A_V$ determination from the host galaxy SED fitting for GRB 010222.


\end{document}